\let\latexaddtocontents\addtocontents
\documentclass[a4paper,twocolumn,11pt,accepted=2026-07-16]{quantumarticle}
\let\addtocontents\latexaddtocontents
\pdfoutput = 1

\usepackage{amssymb,color}
\usepackage{bm} 
\usepackage{color}
\usepackage{graphicx}
\usepackage{physics}
\usepackage{amsthm}
\usepackage{amsmath}
\usepackage{amssymb}
\usepackage{enumerate}
\usepackage{placeins}
\usepackage{booktabs}
\usepackage{dsfont}

\usepackage{microtype}
\usepackage{float}
\usepackage{mathtools}
\usepackage{hyperref}

\newcommand{\e}{{\rm e}}  
\newcommand{\ren}{R\'{e}nyi~}
\newcommand{\renh}{R\'{e}nyi-}

\usepackage[backend=bibtex,url=false]{biblatex}
\begin{document}
\title{Detection of a \ren Index Dependent Transition in Entanglement Entropy Scaling}
\author{Hatem Barghathi}
\affiliation{Department of Physics and Astronomy, University of Tennessee, Knoxville, TN 37996, USA}
\affiliation{Institute for Advanced Materials and Manufacturing, University of Tennessee, Knoxville, Tennessee 37996, USA\looseness=-1}

\author{Adrian Del Maestro}
\affiliation{Department of Physics and Astronomy, University of Tennessee, Knoxville, TN 37996, USA}
\affiliation{Institute for Advanced Materials and Manufacturing, University of Tennessee, Knoxville, Tennessee 37996, USA\looseness=-1}
\affiliation{Min H. Kao Department of Electrical Engineering and Computer Science, University of Tennessee, Knoxville, TN 37996, USA}
	
%\date{\today}

\begin{abstract}
The scaling of entanglement with subsystem size encodes key information about phases and criticality, but the von Neumann entropy is costly to access in experiments and simulations, often requiring full state tomography. The second \ren entropy is readily measured using two-copy protocols and is often used as a proxy for the von Neumann entanglement entropy, where it is assumed to track its asymptotic scaling. Sugino and Korepin (Int. J. Mod. Phys. B 32, 1850306 (2018)) revealed that in the ground state of some highly constrained spin models, the scaling of the von Neumann and \ren entropies can differ, varying from power law to logarithmic scaling as a function of the \ren index.  Here, we construct a number-conserving many-body state that demonstrates a R\'enyi-index-dependent change in the leading entanglement scaling, generalizing previous results to the case of fermionic state. We introduce a symmetry-aware lower bound on the von Neumann entropy built from charge-resolved \ren entropies that can provide a protocol for signaling anomalous entanglement scaling from experimentally accessible data.
\end{abstract}

\maketitle

\section{Introduction}

In a pure quantum state $\vert\Psi\rangle$ of a $d$-dimensional quantum system, the entanglement entropy between a spatial region of size $\ell^d$ and the rest of the system, known as bipartite entanglement, is encoded in the reduced density matrix $\hat{\rho}_A$ associated with partition $A$, which encompasses the degrees of freedom residing locally within it. The entanglement entropy can be quantified using the von Neumann entropy $S_{\rm{vN}}(\hat{\rho}_A)$, or, more generally, by applying the \ren entropy measures $S_\alpha(\hat{\rho}_A)$, indexed by $\alpha$, on the spectrum of the normalized $\hat{\rho}_A$ ($\Tr \hat{\rho}_A = 1$), where for integer values of $\alpha>1$, the \ren entropy can be connected to the expectation value of a swap operator that exchanges the configurations between $\alpha$  replicas of the system within region $A$ \cite{Calabrese:2004ll}.  This connection provides a practical method for estimating the \ren entropy using experimental measurements and computational tools like quantum Monte Carlo (QMC) techniques \cite{Daley:2012rt,Islam:2015ap,Hastings:2010xx,McMinis:2013om,Grover:2013sw}.

On the other hand, the entanglement spectrum and thus all \ren measures of entanglement, including the von Neumann entanglement entropy, can be accessed using density matrix renormalization group (DMRG) methods \cite{Pollmann:2010,Pizorn:2013,Deng:2011}.  Here, the efficient description of the ground state of the system as a matrix product state relies heavily on the scaling of entanglement entropy with system size, where area-law scaling of entanglement is expected to take place \cite{Srednicki:1993,Eisert:2010al,Calabrese:2004ll,Hastings_2007,Anshu:2022}.  The area-law conjecture posits that the dominant scaling of the entanglement entropy of the ground state of a gapped local Hamiltonian is directly proportional to the boundary area of the partition. 

For the purposes of analyzing the asymptotic scaling properties of the entanglement it is useful to remind the reader of the notion of asymptotic compatibility \footnote{The notation $\asymp$ is standard in asymptotic analysis and analytic number theory, denoting equivalence up to a finite multiplicative constant.}.  Two functions $f(x)$ and $g(x)$, that are positive for sufficiently large $x$, are said to be \emph{asymptotically compatible}, denoted by $f(x)\asymp g(x)$, if there exists a constant $c \in (0,\infty)$ such that $\lim_{x\to \infty }{f(x)}/{g(x)}=c$. 
If instead $\lim_{x\to\infty} f(x)/g(x) = 0$ or $\lim_{x\to\infty} f(x)/g(x) = \infty$, we write $f(x)\not\asymp g(x)$.
Thus, for the entanglement area law, $S\asymp \ell^{d-1}$.  While only formally proven in $d=1$ \cite{Hastings_2007}, the more general conjecture holds significance in quantum Hamiltonian complexity theory, with important implications for the feasibility of simulating quantum many-body systems using classical methods \cite{Eisert:2010al,Srednicki:1993,Horodecki:2009,Hastings_2007}. 
In particular, area-law-violating ground states cannot be efficiently represented by tensor networks with small bond dimension. They thus delineate the boundary between quantum systems that are classically tractable and those requiring full quantum resources. Understanding them can help to clarify the frontier of quantum advantage in simulation.  In the absence of an energy gap, as in the case of critical one-dimensional quantum systems governed by conformal field theory, it is well established that the area law is violated, with entanglement entropy growing logarithmically with the size of the subsystem in the thermodynamic limit \cite{Vidal:2003,Calabrese:2004ll,Holzhey:2994,Korepin:2004}. The logarithmic violation of the area law observed in this context may endure in higher dimensions, similar to its manifestation in the ground state of a free Fermi gas \cite{Wolf:2006,Gioev:2006} and the Bose-Einstein condensate of fixed total particle number\cite{Simon:2002,Herdman:2014vd}. This scaling behavior is evident across all \ren entropies with $\alpha>0$, where it is found that $S_{\alpha}\asymp(1+1/\alpha)\ell^{d-1}\ln\ell$ \cite{Leschke:2014}.  In addition strong violations of the area law often arise in systems with critical points \cite{Swingle:2012yd,Huijse:2013}, fracton orders \cite{Ma:2018cu}, or long-range interactions \cite{Nezhadhaghighi:2013ek,Chakraborty:2024ky,Zhao:2025}. These ground states can reveal new ``entanglement phases'' that go beyond the usual symmetry-breaking paradigm, and might underlie exotic materials or non-Fermi liquids.

It has been noted that several $1d$ spin-models exhibit a violation of the area law in their ground states, yet achieving this outcome typically necessitates a certain level of fine-tuning, or the presence of a large local spin value \cite{Irani:2010,Vitagliano:2010}.  A very interesting case demonstrating a significant deviation from the area law has been documented in the ground state of the Motzkin and Fredkin spin models, where the von Neumann entropy scales as $S_{\rm{vN}}\asymp\sqrt{\ell}$, for local spins with $s>1$ \cite{Movassagh:2016,Dell'Anna:2016,Menon:2024,Wang:2025,AdhikariBeach:2021}. Furthermore, Sugino and Korepin demonstrated that the scaling behavior of the ground state entanglement in the Motzkin and Fredkin models is significantly influenced by the \ren index $\alpha$, where it has been shown that for $\alpha<1$ the entanglement entropy scales with the volume of the subsystem as $S_{\alpha<1}\asymp\ell$, while $\alpha>1$ exhibits logarithmic scaling $S_{\alpha>1}\asymp\ln\ell$ \cite{Sugino:2018}. Therefore, it is clear that different measures of entanglement entropy can show diverse scaling with the system size. 

The observation that certain entanglement measures exhibit a pronounced departure from the area law suggests that the corresponding reduced density matrix (RDM) possesses a large number of relevant eigenvalues. This characteristic presents a computational challenge for DMRG methods in accurately representing such highly entangled states efficiently, requiring a priori knowledge and the need to keep a large number of states for convergence that may be prohibitive based on computational resources. In contrast, QMC simulations and experimental determinations of entanglement entropy encounter distinct obstacles. Notably, the efficient estimation of entanglement entropy in these contexts often requires consideration of \ren entropy measures with integer $\alpha\ge 2$. It is often assumed that knowledge of $S_2$ is enough to fully characterize the entanglement entropy, but this approach may fail to fully capture the scaling behavior of entanglement in the presence of a scaling transition in the index $\alpha$. 

In order to improve understanding of the possible origins of entanglement scaling that violate the area law, in this paper, we provide specific examples using entanglement spectra and quantum state ansatz.  We show that the presence of conservation laws and the corresponding symmetry resolution of the RDM allows for direct tuning of individual symmetry sector contributions to the entanglement spectrum.  As a result, we find that for a specific many-body state composed of $N$ particles on $L$ sites, clear $\alpha$-dependent scaling arises: $S_{\alpha > 1} \asymp \ln \ell$, $S_{\alpha = 1} \asymp \sqrt{\ell} \ln \ell$ and $S_{\alpha < 1} \asymp \ell$.  Beyond the example state considered here, we introduce a protocol and diagnostic measure of entanglement that is sensitive to the presence of a scaling transition, which is accessible in both QMC simulations and experimental measurements of the  entanglement entropy.

Throughout this paper we use the phrase ``R\'enyi-index transition'' to denote a change in the \emph{asymptotic} scaling of $S_\alpha(\ell)$ as a function of the entropic \ren index $\alpha$. This should not be confused with a thermodynamic or quantum phase transition of the underlying Hamiltonian: the state is held fixed while $\alpha$ is varied, and for any finite subsystem $S_\alpha$ is a smooth functional of the same reduced density matrix. The origin of the effect is instead replica biasing: changing $\alpha$ reweights the contributions of different symmetry sectors (and, more generally, different parts of the entanglement spectrum), so that distinct large-deviation sectors dominate the replica sum in the thermodynamic limit. When the sector weights and the sector entanglement scale differently with $\ell$, this change of dominance can produce different leading scalings for $\alpha>1$ and $\alpha\le 1$ without implying any singularity in the physical phase diagram.
%%%%%%%%%%%%%%%%%%

\section{\ren entropy and Symmetry resolved entanglement entropy}
For a pure state $\hat{\rho}$ describing a quantum system, we can quantify the entanglement entropy  between partition $A$ of the system and its complement $\bar{A}$ using the \ren entanglement entropy:
\begin{align}
S_{\alpha}=\frac{1}{1-\alpha}\ln\left[\Tr\hat{\rho}_{A}^{\alpha}\right],
\label{eq:Salpha}
\end{align}
where $\hat{\rho}_{A}=\Tr_{\bar{A}}\hat{\rho}$ is the reduced density matrix of the system and $\alpha\ge 0$ is the \ren index. The expression in Eq.~(\ref{eq:Salpha}) captures different entanglement measures, e.g., in the limit $\alpha\to1$ it corresponds to the von Neumann entanglement entropy, where $S_1=-\Tr\hat{\rho}_{A}\ln\hat{\rho}_{A}$ and for $\alpha=1/2$, it defines the logarithmic negativity $\mathcal{E}$  of a pure state $\rho$, where $\mathcal{E}=\ln\Tr\vert\hat{\rho}^{\Gamma_A}\vert$. Here, $\hat{\rho}^{\Gamma_A}$ is the partial transpose of $\hat{\rho}$ with respect to partition $A$ \footnote{Here $\Tr\vert \hat{O}\vert=\Tr\sqrt{O^{\dagger}O^{\phantom{\dagger}}}$ is the trace norm of operator $\hat{O}$}. 

In general the $\alpha$-dependent \ren measures of entanglement entropy yield different values for the same state, and are related to each other through the inequality:
\begin{align} 
    S_{0}\ge S_{\alpha}\ge S_{\alpha^\prime}\ge S_{\infty},
    \label{eq:inquality1}
\end{align}
where $\alpha\le\alpha^\prime$. The above inequality can be proven using Jensen's inequality \cite{jensen:1906}. For a given $\hat{\rho}_{A}$ defined on a finite partition $A$ and with the set of non-zero eigenvalues $\{\lambda_i\}$, the minimum entropy $S_\infty=-\ln\lambda_{\max}$ is defined only by the largest eigenvalue $\lambda_{\max} \equiv \max_i \lambda_i$. However, the Hartley entropy \cite{Hartley:1928} or maximum entropy $S_0=\ln\vert\{\lambda_i\}\vert$ depends only on the number of non-zero eigenvalues of $\hat{\rho}_{A}$, where $\vert\{\lambda_i\}\vert$ is the cardinality of the set $\{\lambda_i\}$. For a maximally entangled state $\vert\{\lambda_i\}\vert=\mathcal{D}_A$ where $\mathcal{D}_A$ is the size of the subspace $A$. Consequently, such bounds on the \ren entropies provides room for distinct scaling with the underling quantum partition properties such as system size, total charge, or total magnetization for different values of $\alpha$.  However, all of the $\alpha$-\ren entropies coincide with each other if all non-zero eigenvalues of $\hat{\rho}_{A}$ are equal, which is true in two extremes of $\hat{\rho}_{A}$: a pure state ($S_\alpha=0$) and a maximally entangled one ($S_\alpha=\ln\mathcal{D}_A$). 

In general, the complexity of an exponentially large reduced-density matrix $\hat{\rho}_{A}$ obstructs the tunability of its spectrum via microscopic system or partition properties.  In contrast, the subjection of the system to global or local conservation laws and symmetries allows for viewing the spectrum of $\hat{\rho}_{A}$ as the union of subsets indexed by a physical observable.  More rigorously, if we assume that the pure state $\hat{\rho}$ of a quantum system conserves some total charge $\hat{Q}=\sum_i\hat{q}_i$ such that $\hat{Q}\hat{\rho}=Q\hat{\rho}$, where $Q$ could represent, e.g., the total magnetization, electric charge, or number of quantum particles in general, and $\hat{q}_i$ is the corresponding local operator for mode $i$, then the corresponding reduced density matrix $\hat{\rho}_{A}$ satisfies $\left[\hat{\rho}_{A},\hat{q}_A\right]=0$,  where $\hat{q}_A=\sum_{i\in A}\hat{q}_i$. As a result, $\hat{\rho}_{A}$ has a block-diagonal structure, where each block corresponds to a charge value $q_A$. This permits the symmetry resolution of $\hat{\rho}_{A}$ as
$\hat{\rho}_{A}=\sum_{q_A}\hat{\mathcal{P}}_{q_A}\hat{\rho}_{A}\hat{\mathcal{P}}_{q_A}$, where $\hat{\mathcal{P}}_{q_A}$ is the projection operator onto the subspace of fixed $q_A$ \cite{Wiseman:2003ei,Goldstein:2018,Horvath:2020,Murciano:2021,Horvath:2022,Arildsen:2025,Monkman:2023,Monkman:2020oe,Castro-Alvaredo:2025,Parez:2022,Barghathi:2018rg,Melko:2016tv,Barghathi:2018rg,Melko:2016tv,Monkman:2023c2}. The probability of $\hat{\rho}_{A}$ having $q_A$ in partition $A$ is given by the trace of the corresponding block as $P_{q_A}=\Tr[\hat{\mathcal{P}}_{q_A}\hat{\rho}_{A}\hat{\mathcal{P}}_{q_A}]$ and the related RDM $\hat{\rho}_{q_A}=\hat{\mathcal{P}}_{q_A}\hat{\rho}_{A}\hat{\mathcal{P}}_{q_A}/P_{q_A}$ has \ren entanglement entropy 
\begin{align}
S_{\alpha}(q_A)=\frac{1}{1-\alpha}\ln\left[\Tr\hat{\rho}_{q_A}^{\alpha}\right]
\label{eq:Sq}.
\end{align}
Thus, we can rewrite Eq.~(\ref{eq:Salpha}) in terms of the probability distribution $\{P_{q_A}\}$ and the corresponding $S_{\alpha}(q_A)$'s as
\begin{align}
    S_{\alpha}=\frac{1}{1-\alpha}\ln\left[\sum_{q_A}\qty(P_{q_A})^\alpha e^{(1-\alpha)S_\alpha(q_A)}\right]
\label{eq:SPq_alpha},
\end{align}
and in the limit $\alpha \to 1$ we have the von Neumann entanglement entropy given by \cite{Klich:2008se,Barghathi:2019db,Kiefer-Emmanouilidis:2020,Kiefer-Emmanouilidis:2020Sp}
\begin{equation}
    S_{1}=\sum_{q_A}P_{q_A} S_1(q_A)+H_1(\{P_{q_A}\}),
\label{eq:SPq_1}
\end{equation}
where $H_1(\{P_{q_A}\})=-\sum P_{q_A}\ln P_{q_A}$ is the Shannon entropy of the probability distribution $\{P_{q_A}\}$.
Before discussing the implication of the above formulas, let's normalize the modified distribution $P_{q_A}^\alpha$ by defining $P_{q_A}^{(\alpha)}=\pqty{P_{q_A}}^\alpha/\sum_{q_A}\pqty{P_{q_A}}^\alpha$, which allows us to rewrite Eq.~(\ref{eq:SPq_alpha}) as
\begin{align}
    S_{\alpha}\!\!=\!\!\frac{1}{1-\alpha}\!\ln\!\!\left[\!\sum_{q_A}\!P_{q_A}^{(\alpha)}\! e^{(\!1-\alpha\!)S_\alpha(q_A)}\!\right]\!\!+\!H_\alpha(\!\{P_{q_A}\}\!).
\label{eq:SPq_Palpha}
\end{align}
Here, $H_\alpha(\{P_{q_A}\})=\frac{1}{1-\alpha}\ln\sum P_{q_A}^\alpha$ is the \ren entropy of the probability distribution $\{P_{q_A}\}$. Thus, the above equations allow us to express $S_{\alpha}$ in terms of the probability distribution $\{P_{q_A}\}$ and the symmetry-resolved entanglement entropies $S_\alpha(q_A)$.
%%%%%%%%%%%%%%%%%%

\section{Entanglement entropy scaling analysis}
\label{sec:EntropyScalingAnalysis}
In this section, starting from the spectrum of the reduced density matrix under the influence of conservation laws, we show how the entanglement entropy can depend on the \ren index $\alpha$, while in the next section we provide an explicit example at the state level. Consider specializing to the case where the general charge $\hat{Q} \equiv \hat{N}$ measures the total number of particles in a many-body state $\hat{\rho}$ describing a quantum system on $L$ lattice sites with $N$ particles (fixed) such that $\hat{N} \hat{\rho} = N \hat{\rho}$, where the filling $N/L$ is fixed. In addition, we consider a bipartition of the system into two spatial subsystems having $\ell$ and $L-\ell$ sites, such that the ratio $\ell/L$ is fixed as $L$ increases, thus both $N$ and $\ell$ scale linearly with $L$. Without loss of generality, we consider a half-partition ($L=2\ell$). For the projected RDM $\hat{\rho}_{\ell,n}=\hat{\mathcal{P}}_{\ell,n}\hat{\rho}_{\ell}\hat{\mathcal{P}}_{\ell,n}/P_{n}$ with $n$ particles in partition $\ell$, the corresponding symmetry-resolved entanglement entropy $S_\alpha(n)$ vanishes for an empty ($n=0$) or fully occupied partition ($n=N$). In contrast, the maximum rank of $\hat{\rho}_{\ell,n}$ is achieved at $n=N/2$ representing a bound on $S_\alpha(n=N/2)\asymp\ell\asymp N/2$ that scales linearly with $\ell$.  

% -------------------------------------------------------------------------------
\subsection{Scaling Examples}
\label{ScalingExamples}
% -------------------------------------------------------------------------------

To gain insight into the possible dependence of $S_\alpha$ on the \ren index $\alpha$, we consider some specific examples.  For simplicity, suppose that for a given number of particles $n$ in partition $\ell$, the entanglement $S_\alpha(n)$ scales linearly with $n$, as $S_\alpha(n)=B\min\{n,N-n\}$. Here we drop the possible dependence of the coefficient $B$ on $\alpha$ by further assuming that the non-vanishing eigenvalues of $\hat{\rho}_{\ell,n}$ are equal (the generalization to non-equal eigenvalues is straightforward). Under these considerations, Eqs.~(\ref{eq:SPq_1}) and (\ref{eq:SPq_alpha}) yield:
\begin{align}
    S_{1}=B\langle \min\{n,N-n\}\rangle_1+H_1(\{P_{n}\})
\label{eq:SPn_1},
\end{align}
and
\begin{align}
    S_{\alpha}\!=\!\!\frac{1}{1-\alpha}\!\ln\langle e^{(\!1-\alpha\!)B\!\min\{n,N-n\}}\rangle_\alpha\!\!+\!\!H_\alpha(\!\{P_{n}\}\!),
\label{eq:SPn_alpha}
\end{align}
where $\langle\dots\rangle_\alpha$ indicates an average over the renormalized probability distribution $\qty{P_n^{(\alpha)}}$. The second term in the above equations represents the entropy of the distribution $\qty{P_n}$, which is bounded from above by the logarithm of the number of possible values $n$ can take, \emph{i.e.}, $H_\alpha(\{P_{n}\})\leq H_0(\{P_{n}\})\leq \ln \left(N+1\right)$. However, the scaling of the first term in Eqs. (\ref{eq:SPn_1}) and (\ref{eq:SPn_alpha}) with the system size $L$ could vary significantly with $\{P_n\}$. For example, for a perfectly peaked distribution at $n=n_0$, all of the entanglement measures coincide with $Bn_0$ and thus share the same scaling of $n_0$ with the system size $L$. In contrast, If we consider $\{P_n\}$ to be flat, such that $P_n=1/(N+1)$ and $N\gg1$, we find
\begin{align}
S_1= B \frac{N}{4}+\ln(N)+\mathcal{O}\qty(\frac{1}{N})
\label{eq:S1_flat},
\end{align}
and
\begin{align}
    S_{\alpha} &=\frac{1}{1-\alpha}\ln\left[\sum_n e^{(1-\alpha)B\min\{n,N-n\}}\right] \nonumber \\
               & \quad+\frac{\alpha}{\alpha-1}\ln(N+1).
\end{align}
For $\alpha<1$, we find a volume law scaling:
\begin{align}
    S_{\alpha<1}\!=\! B\frac{N}{2} \!+\!C_{\alpha(B)}\!+\!\frac{\alpha}{\alpha-1}\ln N\!+\!\mathcal{O}\qty(\frac{1}{N}),
\end{align}
which is consistent with the scenario for $\alpha = 1$ and the constant $C_\alpha$ is defined in Appendix~\ref{app:constants}.

On the other hand, the situation changes when considering $\alpha>1$, as follows:
\begin{equation}
    S_{\alpha>1}\!\!=\!\!\frac{\alpha}{\alpha\!-\!1}\!\ln \!N\!+\frac{1}{1\!-\!\alpha}\!\ln{\frac{2}{1\!-\!e^{(\!1-\alpha\!)B}}}+\mathcal{O}\qty(\!\!\frac{1}{N}\!\!),\!\!
\end{equation}
where the entanglement entropy increases logarithmically with $N$.  Thus, for this specific case we observe a transformation of the scaling of $S_\alpha$ from a volume law to  logarithmic scaling as a result of modifying the \ren index $\alpha$: $S_{\alpha \le 1} \not\asymp S_{\alpha > 1}$. 

\begin{figure}[t!]
    \centering
    \includegraphics[width=8.6cm]{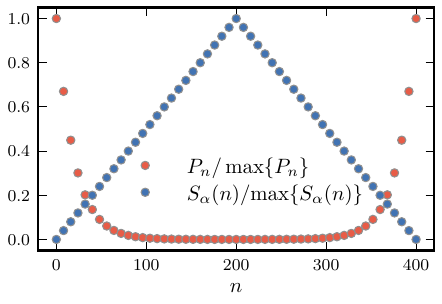}
    \caption{Particle number distribution $P_n$ and symmetry-resolved entanglement entropy $S_\alpha(n)$ for the example discussed in Section \ref{sec:EntropyScalingAnalysis} for a system with $L=800$ sites and $\mu(N)\asymp \sqrt{N}$.}
    \label{fig:PnExample}
\end{figure}

\begin{figure}[t!]
    \centering
    \includegraphics[width=8.6cm]{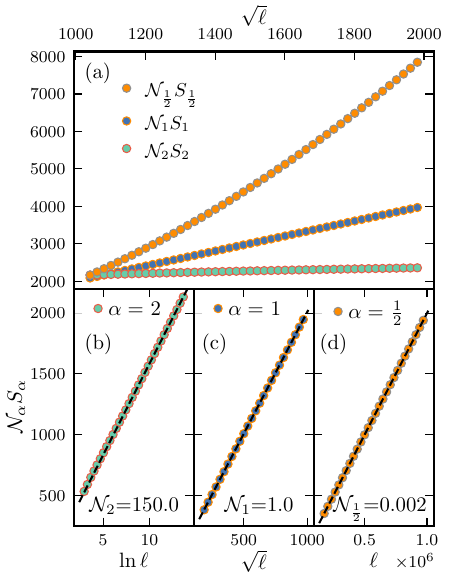}
    \caption{The entanglement spectrum displays diverse entanglement entropy scaling with a partition size of $\ell$. Panel (a) shows the \ren entanglement for $\alpha = 1/2,1,2$ vs. $\sqrt{\ell}$. The asymptotic scaling can be seen more clearly in the lower panels: (b) $S_{\alpha>1} \asymp \ln \ell$, (c) $S_{1} \asymp \sqrt{\ell}$, and (d)  $S_{\alpha<1} \asymp \ell$. Here $\mu(N)\asymp \sqrt{N}$.  The values $\mathcal{N}_\alpha$ are normalization constants chosen to allow a common $y$-axis with their values shown in the lower panels.}
\label{fig:EEScalingS}
\end{figure}

Next we investigate an intermediary exponential distribution that falls between a perfectly localized and flat one:
\begin{align}
P_n=A_N e^{-\min\{n,N-n\}/\mu(N)}
\label{Eq:Pn},
\end{align} 
with $\mu(N)\ll N$ for $N\gg1$ and $A_N$ is a normalization constant defined in Appendix~\ref{app:constants}.  The resulting $P_n$ is shown in Fig.~\ref{fig:PnExample}.
%%%
%%%
In this scenario, and for $N\gg1$, we identify three unique scaling regimes based on the value of $\alpha$ as shown in Fig.~\ref{fig:EEScalingS}.
%%%
%%%

\subsubsection{$\alpha = 1$}
For $\alpha = 1$, we find that the von Neumann entropy is given by
\begin{align}
    S_1=& B\mu(N)+\ln\left[2\mu(N)\right] \nonumber \\
        &-B/2+1+\mathcal{O}\pqty{\frac{1}{\mu(N)}},
\end{align} 
where the leading term of the entanglement entropy scales with the distribution rate parameter $\mu(N)$.  This can be understood via an analysis of Eq. (\ref{eq:SPq_1}), where the exponential decay of $P_n$ in conjunction with a decay constant $\mu(N)$ implies that averaging any quantities with sub-exponential growth rates will be dominated by $\mu(N)$. Thus, we expect that the average value of $S_1(n)$ will be primarily influenced by the scaling behavior of $S_1(n\asymp\mu(N))$.

\subsubsection{$\alpha < 1$}
In contrast, the \ren entanglement entropy with $\alpha<1$ shows a volume law scaling:
\begin{align}
    S_{\alpha<1} & = \frac{BN}{2}-\frac{\alpha}{1-\alpha} \bqty{\frac{N}{2\mu(N)} + \ln 2 \mu(N)} \nonumber \\
                 & \quad + D_\alpha(B)+\mathcal{O}\pqty{\frac{1}{\mu(N)}}.
\label{eq:SIII_alphalt1}
\end{align} 
where $D_\alpha(B)$ is a constant defined in Appendix~\ref{app:constants}.
For $\alpha<1$, the summation in Eq.~\eqref{eq:SPq_alpha} involves terms with exponents $-\alpha \min\{n,N-n\}/\mu(N)+(1-\alpha)S_\alpha(n)$. The dominance of certain terms within the summation is contingent upon the growth rate of $S_\alpha(n)$ relative to $\min\{n,N-n\}/\mu(N)$; specifically, larger values of $S_\alpha(n)$ will prevail if $S_\alpha(n)$ increases at a faster pace than $\min\{n,N-n\}/\mu(N)$. Consequently, it can be anticipated that the predominant terms will correspond to those with higher values of $S_\alpha(n)$, which in this instance, scale proportionally with $N/2$ or equivalently with $\ell$. As a result the entanglement entropy scales with $\ell \asymp N$ as seen in Eq.~\eqref{eq:SIII_alphalt1}.  

\subsubsection{$\alpha > 1$}
Finally, for $\alpha>1$, the \ren entanglement entropy shows even more distinct scaling, where the entanglement grows logarithmically with $\mu(N)$: 
\begin{align}
    S_{\alpha>1} & = \frac{\alpha\ln2\mu(N)}{\alpha-1}+\frac{1}{\alpha-1}\ln\frac{1-e^{-(\alpha-1)B}}{2} \nonumber \\
                 & \quad +\mathcal{O}\pqty{\frac{1}{\mu(N)}}.
\end{align} 
Here, the negative coefficient in the exponent of Eq.~\eqref{eq:SPq_alpha} $-\alpha \min\{n,N-n\}/\mu(N)+(1-\alpha)S_\alpha(n)$ acts to diminish the impact of large $S_\alpha(n)$ values in the overall result. The logarithmic relationship between $S_\alpha(n)$ and $\mu(N)$ stems from the logarithm of both the normalization constant $A_N$ of $P_n$ and its entropy $H_{\alpha>1}(\{P_{n}\})$. 

This simple example serves to elucidate the diverse \renh index dependent scaling behaviors that can be observed for a given entanglement spectrum. This naturally leads to an intriguing question: Is it possible to identify a many-body quantum state which has a strongly \renh index dependent entanglement scaling?  In particular, does the existence of such states extend to systems characterized by a minimal number of local degrees of freedom, including fermions, hardcore bosons, or spin-1/2 systems?
%%%%%%%%%%%%%%%%%%

\section{Example: Many body state}
\label{Sec:ManyBodyStateExample}
To answer this question,  we consider a quantum state describing a system with only two local degrees of freedom on each of $L$ sites, (e.g.\@ fermions or hard-core bosons). For a fixed total number of $N$ particles, the many-body state can be written as:
\begin{align}
\vert\Psi\rangle=\sum_n\sqrt{P_n}\vert\psi_{n,N-n}\rangle
\label{Eq:StateExample},
\end{align} 
where $\vert\psi_{n,N-n}\rangle$ represents the projection of the state $\vert\Psi\rangle$ on a fixed number of particles $n$ in the partition of $\ell=L/2$ sites and $N-n$ particles in the complementary partition. Here, $P_n$ is the corresponding particle number distribution.

\begin{figure}[t!]
    \centering
    \includegraphics[width=8.6cm]{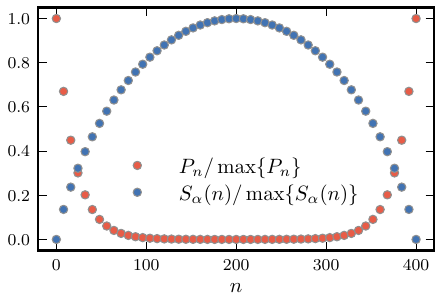}
    \caption{Particle number distribution $P_n$ and symmetry-resolved entanglement entropy $S_\alpha(n)$ analysis using the quantum state $\vert\Psi\rangle$ introduced in Eq.~(\ref{Eq:StateExample}). Here, the state $\vert\Psi\rangle$ describes $N=400$ particles with $\mu(N)\asymp \sqrt{N}$.}
    \label{fig:PnExampleState}
\end{figure}

Guided by intuition gained from understanding symmetry resolved entanglement,  we seek to maximally entangle the states $\vert\psi_{n,N-n}\rangle$.  This can be achieved by pairing-up sites in both partitions such that if a site in region $A$ is occupied, its complement should be empty (or vice versa).  Accordingly, we take the ansatz
\begin{align}
    \vert\psi_{n,N-n}\rangle\!=\!{\frac{1}{\sqrt{{\ell}\choose{n}}}}\!\sum_{\qty{n_i}}\!\vert n_1,\dots, n_\ell,\bar{n}_1,\dots, \bar{n}_\ell\rangle,
    \label{eq:PsinN}
\end{align} 
where $\bar{n}_i=1-n_i$ and the summation run over all of the ${{\ell}\choose{n}}$  possible configurations $\{n_1,\dots, n_\ell\}$ of the $n=\sum_{i=1}^\ell n_i$ particles. Eq.~\eqref{eq:PsinN} can be seen to already be in the form of a Schmidt decomposition: 
\begin{align}
\vert\psi_{n,N-n}\rangle\!=\!\!\sum_{\qty{n_i}}\!\!{\frac{1}{\sqrt{{\ell}\choose{n}}}}\vert n_1,\dots, n_\ell\rangle_{\ell}\!\otimes\!\vert\bar{n}_1,\dots, \bar{n}_\ell\rangle_{L-\ell},
\end{align} 
where $\vert n_1,\dots, n_\ell,\bar{n}_1,\dots, \bar{n}_\ell\rangle=\vert n_1,\dots, n_\ell\rangle_{\ell}\otimes\vert\bar{n}_1,\dots, \bar{n}_\ell\rangle_{L-\ell}$. The equal ${{\ell}\choose{n}}$ Schmidt coefficients ensures maximal symmetry resolved entanglement 
\begin{align}
S_\alpha(n;\ell)=\ln{{\ell}\choose{n}}.
\label{Eq:Sn}
\end{align} 
These states can then be paired with the exponentially decaying probability distribution introduced in Eq.~(\ref{Eq:Pn}) to construct the full many-body state with the results shown in Figure~\ref{fig:PnExampleState} with $\mu = \sqrt{N}$. 
%%%
%%%

\begin{figure}[t]
\centering
\includegraphics[width=8.6cm]{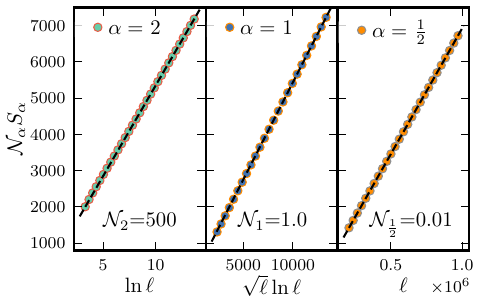}
\caption{Different measures of entanglement entropy applied to the state $\vert\Psi\rangle$ defined in Eq.~(\ref{Eq:StateExample}) for  $\mu(N)\asymp \sqrt{N}$. Asymptotically, $S_{\alpha = 2} \asymp \ln \ell$ (left), $S_1 \asymp \sqrt{\ell}\ln\ell$ (middle), and  $S_{\alpha=1/2<1} \asymp \ell$ (right).  The values $\mathcal{N}_\alpha$ are normalization constants chosen to allow a common $y$-axis.}
\label{fig:EEScaling}
\end{figure}

Numerically calculating the entanglement entropy for different values of $\alpha$, confirms distinct scaling with the size of the subregion $\ell$ 
for $\alpha=2$, $\alpha=1$, and $\alpha=1/2$, as demonstrated in Fig.~\ref{fig:EEScaling} where it is clear that $S_{1/2} \not\asymp S_1 \not\asymp S_2$.  
%%%
%%%

In analogy with the analysis presented in Section~\ref{ScalingExamples}, for $\alpha=1/2<1$, the entanglement exhibits volume law scaling proportional to $\ln{{\ell}\choose{n\approx N/2}}\asymp\ell$, where  $\ell=N$ for half-filling and a half-partition. Conversely, for $\alpha=2>1$, the entanglement displays the anticipated logarithmic scaling with $\ell$. However, in the $\alpha=1$ case, we observe a multiplicative logarithmic correction to the previously found $\sqrt{\ell}$ scaling. This can be understood (full details are included Appendix~\ref{app:log}) by examining the scaling properties of $\ln{{\ell}\choose{n}}$, with $n\asymp\mu(N=\ell)$, where for $n\asymp\mu(N)\asymp N^{\gamma<1}$ we have $\ln{{\ell}\choose{\mu(N=\ell)}}\asymp\ell^\gamma\ln \ell$. 

Thus, we have identified an example of a quantum state describing a limited number of local degrees of freedom (specifically, two) that showcases a transition in the entanglement entropy scaling behavior for a half partition from volume law to logarithmic scaling as a function of the \ren index $\alpha$. We note that the entanglement scaling transition demonstrated in Fig.~\ref{fig:EEScaling} is not unique to this specific example which has maximal symmetry resolved entanglement. For example, a state with less symmetry resolved entanglement, $S_\alpha(n;\ell) = \sqrt{\ln{\ell}\choose{n}}$, can be shown to have $S_1(\ell) \asymp \ell^{1/4}\sqrt{\ln\ell}$.

In the next section we propose a useful measure for the detection of such states.
%%%%%%%%%%%%%%%%%%

\section{Comparison with the Motzkin Spin Chain}
\label{sec:MotzkinSpinChain}

In this section, we draw a direct comparison between the entanglement spectrum of our $U(1)$-symmetric states and that of Motzkin spin-chain ground states. Despite arising from distinct physical frameworks, we demonstrate that both exhibit similar spectral structure, which gives rise to the anomalous sub-volume scaling of the entanglement entropy. The Motzkin spin-$s$ chain provides a useful point of comparison, as it is one of the few models for which the full entanglement spectrum of the reduced density matrix (RDM) is known analytically. For a half-chain partition ($\ell = L/2$) \cite{Movassagh:2016,Dell'Anna:2016,Menon:2024,Wang:2025, AdhikariBeach:2021,Sugino:2018}, the entanglement spectrum is organized by the height $h \geq 0$ of the Motzkin walk at the midpoint of the chain. For each value of $h$, there are $s^h$ degenerate eigenvalues of the RDM, each equal to $P_h / s^h$, where $P_h$ is the total weight carried by height-$h$ configurations. In the asymptotic limit of large $\ell$, the weights $P_h$ take the form of a discretized half-Gaussian
\cite{Sugino:2018,Menon:2024},
\begin{align}
    P_h \sim \sqrt{\frac{2}{\pi\sigma^3}}
    \frac{(h+1)^2}{\ell^{3/2}}
    \exp\!\left(-\frac{(h+1)^2}{2\sigma\ell}\right),
    \label{Eq:MotzkinEigenvalues}
\end{align}
where the parameter $\sigma = \sqrt{s}/(1 + 2\sqrt{s})$ encodes the spin magnitude $s$ and controls the width of the distribution over $h$. For $s = 1$, Eq.~\eqref{Eq:MotzkinEigenvalues} reduces to the standard Motzkin chain result, while larger $s$ enhances the entanglement. The normalization condition $\sum_{h=0}^{\ell} P_h = 1$ is satisfied asymptotically.

The entanglement structure of the Motzkin chain admits a natural two-level description: within each height sector $h$, the $s^h$ degenerate eigenvalues contribute an intra-sector entropy $S_\alpha(h) = h \ln s$, while the distribution $\{P_h\}$ over height sectors contributes an inter-sector entropy.

The total \ren entanglement entropy follows from the two-level
decomposition as
\begin{align}
    S_{\alpha} = \frac{1}{1-\alpha}
    \ln\!\left[
        \sum_{h=0}^{\ell}
        \left(P_h\right)^\alpha
        e^{(1-\alpha)S_\alpha(h)}
    \right],
    \label{eq:Sh_alpha}
\end{align}
where the factor $e^{(1-\alpha)S_\alpha(h)} = s^{(1-\alpha)h}$
accounts for the degeneracy within each height sector.
In the von Neumann limit $\alpha \to 1$, Eq.~\eqref{eq:Sh_alpha}
reduces to
\begin{equation}
    S_{1} = \sum_{h=0}^{\ell} P_h \, S_1(h) + H_1(\{P_h\}),
    \label{eq:Sh_1}
\end{equation}
where $H_1(\{P_h\}) = -\sum_h P_h \ln P_h$ is the Shannon entropy
of the height distribution. 
Using Eq.~\eqref{Eq:MotzkinEigenvalues}, the mean height scales
as $\langle h \rangle \asymp \sqrt{\sigma \ell}$ for large $\ell$,
so the intra-sector contribution scales as
$\langle h \rangle \ln s \asymp\sqrt{\ell} \ln s$,
consistent with the known area-law violation and
$\sqrt{L}$ scaling of entanglement entropy in the Motzkin
chain~\cite{Movassagh:2016,Dell'Anna:2016}. $ S_{\alpha>1}$ and $ S_{\alpha<1}$ scale like $\ln L$ and $L$, respectively \cite{Sugino:2018}.
%%%%%%%%%%%%%%%%%%

\section{Detecting the Entanglement Transition}
\label{sec:proposedMeasures}

The \ren entanglement entropy with an integer index $\alpha \ge 2$ can be measured in quantum simulations and ultra-cold atomic experimental setups by employing $\alpha$ (usually two) replica copies of the system \cite{Calabrese:2004ll, Daley:2012rt,Islam:2015ap,Kaufman:2016rq,Pichler:2016rd, Lukin:2019dy, Brydges:2019zv}. This obviates the need for full state tomography and allows the \ren entropy to be measured via the expectation value of a local \emph{swap} estimator.  Its use has been widespread and allowed for studies of entanglement scaling laws in strongly interacting quantum systems via quantum Monte Carlo \cite{Hastings:2010xx,Isakov:2011yg,Helmes2014,Wang:2014dy,Herdman:2017ts,Wildeboer:2017fj,Zhao:2022aj,Demidio:2024xv,Laflorencie:2016}. However, while it is in principle possible to measure higher \ren entropies and extrapolate towards the von Neumann case of $\alpha=1$, in practice this is computationally prohibitive and most studies consider only the single case of $\alpha=2$, using the $S_2$ scaling as a more general (approximate) proxy for entanglement entropy. Thus we seek a diagnostic protocol that can signal the possibility of a transition in the scaling of entanglement entropy as a function of $\alpha$ that only relies on knowledge of $\alpha = 2$.

This is possible when conservation laws, such as the preservation of total charge $\hat{Q}=\sum_i\hat{q}_i$ are present, where the symmetry resolution of the \ren entanglement entropy $S_{\alpha}(q_A)$ for $\alpha>1$, along with the associated distribution of local charge $\{P_{q_A}\}$ \cite{CasianoDiaz:2023pi} can be calculated. In practice, a histogramming technique can be employed to resolve the \ren entanglement in conjunction with a projective measurement of the observable $q_A$. 

We posit the existence of the desired measure, which we label $\bar{S}_\alpha$, that is sensitive to the asymptotic scaling of the von Neumann entanglement entropy.  For this quantity to be useful, we strictly require that 
\begin{align}
\bar{S}_{\alpha}=\mathcal{O}(S_1),
\label{Eq:Sn1}
\end{align} 
such that the scaling of $\bar{S}_{\alpha}$ with system size cannot exceed that of $S_1$. Also, it is favorable that  
\begin{align}
S_{\alpha}=\mathcal{O}(\bar{S}_{\alpha}),
\label{Eq:Sn2}
\end{align} 
such that scaling $\bar{S}_{\alpha}$ is not weaker than the scaling of $S_{\alpha}$ \footnote{For any two positive functions  $f(x)$ and $g(x)$, $f=\mathcal{O}(g)$ is defined as: $\exists$ $x_0$ and a constant $c > 0$ such that  $\forall$  $x > x_0$, $f(x) \leq cg(x)$.}.

These conditions establish a notably lenient framework for the measure $\bar{S}_{\alpha}$. Nevertheless, the primary objective of such a measure is to address the issue of attenuating the impact of large $S_\alpha(q_A)$ values as a result of the negative factor $1-\alpha$ present in the $e^{(1-\alpha)S_\alpha(q_A)}$ terms in Eq.~\eqref{eq:SPq_alpha}. To better elucidate this concept, we first rewrite Eq.~\eqref{eq:SPq_alpha} as 
\begin{align}
    S_{\alpha}=-\ln\left[\sum_{q_A}P_{q_A}X_{q_A}^p\right]^{\frac{1}{p}}
\label{eq:Savg1},
\end{align}
where $S_{\alpha}$ can now be interpreted as the negative logarithm of a power mean $\bar{X}_p=\left[\sum_{q_A}P_{q_A}X_{q_A}^p\right]^{\frac{1}{p}}$ of the non-negative entries $X_{q_A}=P_{q_a}e^{-S_\alpha(q_A)}$, with the positive power  $p=\alpha-1$. Using the inequality $ \bar{X}_0\leq  \bar{X}_p$, where $ \bar{X}_0=\prod_{q_A}X_{q_A}^{P_{q_A}}$ is the geometric mean,  we can write 
\begin{align}
    S_{\alpha}\leq-\ln\prod_{q_A}X_{q_A}^{P_{q_A}}
\label{eq:Savg2}
\end{align}
which can be expanded as
\begin{align}
S_{\alpha}\leq\sum_{q_A}P_{q_A}S_\alpha(q_A)+H_1(\{P_{q_A}\})\, . 
\label{eq:Savg22}
\end{align}
Alternatively, one can use the Jensen’s inequality to prove the last inequality (see Appendix B).  Also, in view of Eq.~\eqref{eq:SPq_1} and the inequality in Eq.~\eqref{eq:inquality1} we see that
\begin{align}
S_{1}\geq\sum_{q_A}P_{q_A}S_\alpha(q_A)+H_1(\{P_{q_A}\})\, .
\label{eq:Savg23}
\end{align}
Given the preceding two inequalities, the sought after definition of $\bar{S}_\alpha$ is now evident
\begin{align}
\bar{S}_{\alpha}\triangleq\sum_{q_A}P_{q_A}S_\alpha(q_A)+H_1(\{P_{q_A}\})
\label{eq:Savg3},
\end{align}
which satisfies the inequalities
\begin{align}
    S_1\geq\bar{S}_{\alpha}\geq S_{\alpha} \qquad \alpha > 1
\label{eq:Savg4}.
\end{align}
This guarantees that $\bar{S}_{\alpha}=\mathcal{O} (S_1)$ and $S_{\alpha}=\mathcal{O}(\bar{S}_{\alpha})$ as originally required. 

To appreciate the utility of the newly proposed measure $\bar{S}_{\alpha}$, we return to the aforementioned scenario where results from an experiment or quantum Monte Carlo simulation have determined $S_{\alpha}$ along with its associated symmetry resolution $S_{\alpha}(q_A)$, alongside the probability distribution $P_{q_A}$, for different system sizes and limited to $\alpha\ge2$. The available information is then sufficient to calculate the corresponding values of $\bar{S}_{\alpha}$ and assess its scaling with respect to $\ell$, and compare it to that of $S_{\alpha}$. The goal of such a comparison is to deduce whether or not $S_{\alpha}$ and $\bar{S}_{\alpha}$ are asymptotically compatible. 

Let's consider the first scenario where $S_{\alpha}$ and $\bar{S}_{\alpha}$ are not asymptotically compatible, denoted as $\bar{S}_{\alpha}\not\asymp S_{\alpha}$, then it follows from Eq.~\eqref{eq:Savg4} that $S_{\alpha} = {o}(\bar{S}_{\alpha})$ or equivalently $\lim_{\ell\to\infty}{S_{\alpha}}/{\bar{S}_{\alpha}}=0$. Given that $\bar{S}_{\alpha}=\mathcal{O} (S_1)$, or equivalently $\lim_{\ell\to\infty}{\bar{S}_{\alpha}}/{S_{1}}=c\in [0,\infty)$ as is granted by Eq.~\eqref{eq:Savg4}, we can write
\begin{align}
\lim_{\ell\to\infty}\frac{S_{\alpha}}{S_{1}}\!=\!\left(\lim_{\ell\to\infty}\frac{S_{\alpha}}{\bar{S}_{\alpha}}\right)\!\!\left(\lim_{\ell\to\infty}\frac{\bar{S}_{\alpha}}{S_{1}}\right)\!=\!0\!\times \!c\!=\!0,
\end{align}
where we have used the fact that the limit of a product can be expressed as the product of non-infinite limits. Thus we have shown that for the case of interest: 
\begin{equation}
S_{\alpha} \not\asymp \bar{S}_{\alpha} \Rightarrow  S_{\alpha} \not\asymp S_1 \qquad
\alpha\ge 2
\label{eq:Sfinal}
\end{equation}
confirms the presence  of an entanglement scaling transition between $\alpha=1$ and 2.

\begin{figure}[t!]
    \centering
    \includegraphics[width=8.6cm]{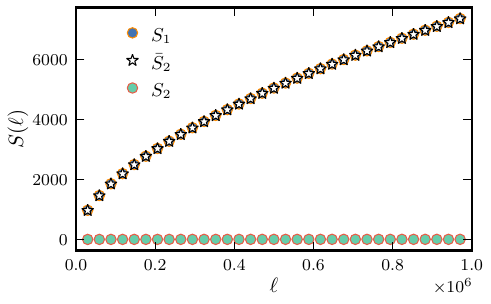}
    \caption{The von Neumann entanglement entropy $S_1$ shows a distinct scaling from that of the \ren entanglement entropy $S_2$, while $\bar{S}_2$ exhibits compatible scaling with $S_1$ for the many-body state considered in section~\ref{Sec:ManyBodyStateExample}.}
    \label{fig:EEAverage}
\end{figure}

Accordingly, the defined measure $\bar{S}_{\alpha}$ can now be used to identify a refined lower bound on $S_1$ with greater sensitivity to the scaling of the symmetry resolved entanglement $S_{\alpha}(q_A)$ than $S_{\alpha>1}$ alone. 
This can be seen in Fig.~\ref{fig:EEAverage}
%%%
%%%
where we demonstrate the similar scaling of $S_1$ and $\bar{S}_2$ with the subsystem size $\ell$ for the example considered in section~\ref{Sec:ManyBodyStateExample}, where  $S_1\sim\bar{S}_2\asymp \sqrt{\ell}\ln{\ell}$, while $S_2\asymp\ln \ell$. For the particular many-body state considered here, the proposed bound $\bar{S}_{\alpha}$ demonstrates an asymptotic growth rate that exceeds the corresponding \ren entanglement entropy $S_{\alpha}$.  Consequently, it can detect the previously found $\alpha$-dependent transition in the entanglement entropy scaling.

We have shown that $\bar{S}_{\alpha}$ can be a useful diagnostic to detect anomalous entanglement scaling, however there is unfortunately no guarantee that it will always exhibit distinct scaling with $\ell$ compared to $S_{\alpha}$. In the scenario where  $\bar{S}_{\alpha}\asymp S_{\alpha}$, the inequalities in Eq.~\eqref{eq:Savg4} provide no useful information about the scaling of $S_1$.  However, if  $S_{1}$ is known and $S_1 \asymp S_{\alpha}$, then it follows that  ${S}_{1}\asymp \bar{S}_{\alpha}\asymp S_{\alpha}$.  

This behavior can be observed in the symmetry-resolved entanglement of a one-dimensional fermionic system described by the Tomonaga-Luttinger model \cite{Giamarchi:2004bk}. Here, the asymptotic particle number distribution in subsystem $\ell$ is Gaussian 
\begin{align}
   P_n\approx \sqrt{\frac{\pi}{2K\ln \ell}}e^{-\frac{\pi^2(\Delta n)^2}{2K\ln \ell}}
\label{eq:PnLL},
\end{align}
where $K$ is the Luttinger parameter and $\Delta n=n-\bar{n}$ and $\approx$ indicates that we have dropped subleading terms in $\ell$. Here $\bar{n}$ is the average number of particles in the partition of size $\ell$ and the corresponding $H_1(\{P_n\})=-\sum_nP_n\ln P_n$ is
\begin{align}
H_1(\{P_n\})\approx\ln(\sqrt{\frac{2K\ln \ell}{\pi}})+\frac{1}{2}\label{eq:H1LL}.
\end{align}
Goldstein and Sela determined the relation between the symmetry-resolved entanglement $S_\alpha(n)$ and the entanglement entropy $S_\alpha$ as \cite{Goldstein:2018}:
\begin{align}
   (P_n)^\alpha\e^{\left(1-\alpha\right)S_\alpha\left(n\right)}\approx \e^{\left(1-\alpha\right)S_\alpha}\sqrt{\frac{\alpha\pi}{2K\ln \ell}}\e^{-\frac{\alpha\pi^2(\Delta n)^2}{2K\ln \ell}}
\label{eq:SnPnLL}.
\end{align}
Taking the natural logarithm of both sides of Eq.~\eqref{eq:SnPnLL}, averaging over $P_n$  and simplifying via Eq.~\eqref{eq:H1LL} yields
\begin{align}
  \sum_nP_n S_\alpha\left(n\right)\approx& S_\alpha-H_1(\{P_n\})\nonumber\\&+\frac{\alpha-1-\ln\alpha}{2\left(\alpha-1\right)}\, .
\label{eq:SnPnLLlog}
\end{align}
Thus we find
\begin{align}
    \bar{S}_\alpha\approx S_\alpha+\frac{\alpha-1-\ln{\alpha}}{2\left(\alpha-1\right)} \quad \Rightarrow \quad S_\alpha \asymp \bar{S}_\alpha\, .
\label{eq:SbarLL}
\end{align}
For this case, it is known from conformal field theory \cite{Calabrese:2004ll} that $S_1\asymp S_\alpha \asymp\ln\ell$ and thus $\bar{S}_\alpha$ exhibits a similar scaling to $S_\alpha$ and $S_1$. 
 %%%%%%%%%%%%%%%%%%

\section{Discussion} 

This work was motivated by the fact that the entanglement quantity of primary theoretical interest for pure states is the von Neumann entropy $S_1$, while in experiments and in quantum Monte Carlo (QMC) the readily accessible quantity is typically a single integer \ren entropy, most often $S_2$ \cite{Daley:2012rt,Islam:2015ap,Hastings:2010xx}.  In many familiar settings $S_2$ and $S_1$ share the same leading scaling with subsystem size, which has encouraged using $S_2$ as a proxy for $S_1$.  However, the ground states of the Motzkin and Fredkin spin chains provide a clear counterexample: the dominant scaling of $S_\alpha$ changes qualitatively as a function of the \ren index, with a sharp crossover near $\alpha=1$ \cite{Sugino:2018}.  Our results demonstrate that such ``\renh index-dependent'' scaling is not an isolated curiosity of highly constrained spin models: once a conservation law is present, it can arise from a simple and physically transparent mechanism. While we restrict our examples to one dimension for clarity of presentation, the results admit a straightforward generalization to higher dimensions. In the spectral analysis, replacing $\ell$ and $L$ with $\ell^d$ and $L^d$,respectively, preserves the entanglement scaling transition in $d$-dimensions. Similarly, the explicit state construction can be extended to higher-dimensional geometries in a natural way. We emphasize that the transition is driven by the \ren index weighting different charge sectors differently, rather than by spatial structure or locality constraints, and as such does not rely on any feature specific to one dimension.

With a conserved charge, the reduced density matrix $\hat{\rho}_A$ is block diagonal, and Eq.~\eqref{eq:SPq_alpha} expresses the total \ren entanglement entropy as a nonlinear combination of the symmetry-resolved entropies $S_\alpha(q_A)$ and the charge distribution $P_{q_A}$.  The dependence on $\alpha$ enters through the weights $P_{q_A}^\alpha$: for $\alpha>1$ these weights suppress rare sectors, while for $\alpha<1$ they enhance them.  Equivalently, $S_\alpha$ is controlled by a \emph{tilted} distribution $P_{q_A}^{(\alpha)}\propto P_{q_A}^\alpha$ in Eq.~\eqref{eq:Savg4}.  This bias implies that a state can hide substantial entanglement in rare symmetry sectors: it will be essentially invisible to $S_{\alpha>1}$ but can dominate $S_{\alpha\le 1}$.  Our explicit number-conserving construction in Sec.~\ref{Sec:ManyBodyStateExample} demonstrates this scenario concretely: by combining (i) sectors whose internal entanglement scales as $\ln\binom{\ell}{q_A}$ with (ii) a distribution $P_{q_A}$ whose width grows subextensively, we obtain $S_{\alpha>1}\asymp\ln\ell$ while $S_1$ is parametrically larger, $S_1\asymp \ell^{1/2}\ln\ell$ (and becomes extensive for $\alpha<1$).  In this sense the ``transition'' in the leading scaling of $S_\alpha$ reflects a change in which charge sectors dominate the replica sum, rather than a thermodynamic phase transition.

From a practical standpoint, this structure suggests a diagnostic that is compatible with the measurement constraints of experiments and QMC.  The introduced measure $\bar{S}$ in Eq.~\eqref{eq:Savg3} satisfies the bound in Eq.~\eqref{eq:Savg4}, $S_1\ge \bar{S}_\alpha \ge S_\alpha$ for $\alpha>1$.  For $\alpha=2$ all required ingredients are accessible within the same replica-based protocols used to measure $S_2$ \cite{Daley:2012rt,Islam:2015ap,Hastings:2010xx}.  In practice one (i) measures $P_{q_A}$ by histogramming projective measurements of $q_A$ in region $A$; (ii) estimates the sector swap expectation values to obtain $S_2(q_A)$; and (iii) forms $\bar{S}_2$ by adding the Shannon term $H_1$.  Randomized measurement protocols provide an alternative route to $S_2$ (and related sector-resolved moments) without preparing multiple copies \cite{Elben:2018ec,Elben:2019tb,Brydges:2019zv}.  A key point is that $\bar{S}_2(\ell)$ and $S_2(\ell)$ can be compared directly as functions of subsystem size.  If $\bar{S}_2$ and $S_2$ are not asymptotically comparable, $\bar{S}_2\not\asymp S_2$, then $S_2$ cannot be asymptotically comparable to $S_1$ either.  In other words, observing a mismatch between $\bar{S}_2$ and $S_2$ is a direct, symmetry-aware indication that $S_2$ underestimates the scaling of the von Neumann entanglement.

The converse does not hold.  As illustrated by the Tomonaga--Luttinger liquid, $\bar{S}_\alpha$ can track $S_\alpha$ even when $S_1$ differs from $S_\alpha$ by subleading corrections.  Thus $\bar{S}_\alpha\asymp S_\alpha$ is consistent with both ``no transition'' and ``transition masked by finite size'' scenarios, and should be interpreted as inconclusive.  Even in this regime, however, $\bar{S}_2$ remains a rigorous (and often tight) lower bound on $S_1$ that can be computed at essentially no extra conceptual cost once symmetry-resolved data are available.

Beyond the explicit construction analyzed here, it would be valuable to identify additional symmetry-conserving families of states, and, more importantly, microscopic local Hamiltonians whose low-lying states realize them, that exhibit a R\'enyi-index-dependent change in the leading entanglement scaling.  
A recent proposal along these lines suggests that fine-tuning of interactions in Rydberg quantum simulators \cite{mukherjee:2026qs} may realize Motzkin physics with $s\ge1$.

A concrete motivation to search for such behavior comes from measurement-based quantum computation (MBQC), where a computation is driven by adaptive local measurements on a pre-prepared entangled resource state (with the cluster state as the canonical example) \cite{RaussendorfBriegel2001,BriegelEtAl2009MBQC}.  In this setting, entanglement is necessary to evade efficient classical simulation, but the relevant requirement is not simply a large bipartite entropy: universal resources must possess an entanglement structure that grows with system size (as quantified, for example, by entanglement-width measures), and generic highly entangled states can in fact be poor resources \cite{VidalSim2003,VanDenNest2006,GrossFlammiaEisert2009}.  A parallel line of work aims to realize MBQC resources as gapped ground states of local Hamiltonians (and in favorable cases within symmetry- or topology-protected phases) in order to leverage spectral gaps and global constraints for robustness \cite{BrennenMiyake2008,RaussendorfHarrington2007,ElseBartlettDoherty2012}.  Because such Hamiltonian settings often come with conservation laws by design or by microscopic constraint, our symmetry-aware lower bound $\bar{S}_2$ provides a practical way to certify growing von Neumann entanglement from experimentally/QMC-accessible second-\ren data, mitigating cases where $S_2$ underestimates $S_1$ by suppressing rare symmetry sectors.  

Finally, while we have focused on a single conserved $U(1)$ charge, the formalism extends naturally to other Abelian symmetries, and suggests analogous diagnostics for non-Abelian symmetry resolution and gauge constraints, where the sector structure of $\hat{\rho}_A$ is even more constrained \cite{Calabrese:2021sr,Bianchi:2024na}.
%%%%%%%%%%%%%%%%%%

\section{Code and Data Availability}
All data and analysis scripts supporting the findings of this paper are available online \cite{paperrepo}.

\section*{Author Contributions}
H.B. and A.D.M. jointly conceived the project and developed the theoretical framework. H.B. derived the analytical results and carried out the numerical calculations. Both authors analyzed and interpreted the results and contributed to writing and revising the manuscript.

\acknowledgments
We thank R. Mukarjhee and R. Islam for interesting discussions.  The authors acknowledge the support of the U.S.  Department of Energy, Office of Basic Energy Sciences under grant No.~DE-SC0022311.

\appendix

\section{Constants Appearing in Section~\ref{sec:EntropyScalingAnalysis}}
\label{app:constants}
\begin{align*}
C_{\alpha(B)} & =\frac{1}{1-\alpha}\ln\left[\frac{e^{(1-\alpha)B}+1}{e^{(1-\alpha)B}-1}\right] \qquad N~\text{even} \\
C_{\alpha(B)} &=\frac{1}{1-\alpha}\ln\left[\frac{2e^{(1-\alpha)B/2}}{e^{(1-\alpha)B}-1}\right] \qquad N~\text{odd}. 
\end{align*}
\begin{alignat*}{2}
    A_N &= \frac{1-e^{-1/\mu(N)}}{2-2e^{-(\frac{N+1}{2})/\mu(N)}}\qquad & N~\text{even} \\ 
A_N
        &=\frac{1-e^{-1/\mu(N)}}{2-e^{-(\frac{N}{2})/\mu(N)}-e^{-(\frac{N}{2}+1)/\mu(N)}} \qquad &N~\text{odd}.
\end{alignat*}
\begin{alignat*}{2}
    D_\alpha(B) & =\frac{1}{1-\alpha}\ln\frac{3e^{(1-\alpha)B}-1}{e^{(1-\alpha)B}-1}\qquad & N~\text{even} \\ 
    D_\alpha(B) & =\frac{1}{1-\alpha}\ln\frac{2}{e^{(1-\alpha)B}-1}\qquad & N~\text{odd}.
\end{alignat*}
%
%%%%%%%%%%%%%%%%%%

\section{Logarithmic Correction for $\alpha = 1$}
\label{app:log}

In this appendix, we focus on the origin of the logarithmic correction factor observed in the scaling of the von Neumann entanglement entropy when considering the many-body quantum state in Section \ref{Sec:ManyBodyStateExample} (Fig.~\ref{fig:EEScaling}). 

Consider the state in Eq.~(\ref{Eq:StateExample}) at half-partition ($L=2\ell$) and fixing $L=2N$ such that $N=\ell$, with the symmetry-resolved entanglement entropy given by 
\begin{align}
S_\alpha(n;\ell)=\ln{{\ell}\choose{n}},
\label{Eq:Sn_appendix}
\end{align} 
with particle number distribution
\begin{align}
P_n=A_\ell e^{-\min\{n,\ell-n\}/\mu(\ell)}
\label{Eq:Pn_appendix},
\end{align} 
where $\mu\asymp \ell^{\gamma}$ with $\gamma<1$ and $A_\ell$ is the normalization constant. Here, both of $S_\alpha(n;\ell)$ and $P_n$ are symmetrical with respect to exchanging $n$ with $\ell-n$ . The targeted quantity is $\sum_{n=0}^\ell P_nS_1(n;\ell)$
\begin{align}
\sum_{n=0}^\ell P_nS_1(n;\ell)=\sum_{n=0}^\ell A_\ell e^{-\min\{n,\ell-n\}/\mu(\ell)}\ln{{\ell}\choose{n}}
\label{Eq:Savg_appendix}.
\end{align} 
In the limit $\ell\gg n\gg 1$ Stirling's asymptotic approximation for factorials $n!\approx \sqrt{2 \pi n}e^{-n}n^n$ yields:
\begin{align}
    \ln{{\ell}\choose{n}} & \approx n\ln\frac{\ell}{n}-\ln\sqrt{2\pi n}\nonumber \\
                          &\quad+(\ell-n+\frac{1}{2})\ln(1-\frac{n}{\ell})  \nonumber \\
                          & \approx n\ln\frac{\ell}{n}-\ln\sqrt{2\pi n}\nonumber \\
                          & \quad-(\ell-n+\frac{1}{2})\sum_{k=1}^{\infty}\frac{n^k}{k\ell^k}
\label{Eq:Sn_1_appendix}.
\end{align} 
where in the second line we have used the Taylor expansion of $\ln(1-x)=-\sum_{k=1}^{\infty}\frac{x^k}{k}$, where $x=\frac{n}{\ell}\ll 1$.
The exponentially decaying probability distribution in Eq.~(\ref{Eq:Savg_appendix}) will suppress the contribution from terms with $\mu\ll n \ll \ell-\mu $, and the dominant contribution to the summation in Eq.~(\ref{Eq:Savg_appendix}) will come from terms in the vicinity of $\mu$ and $\ell-\mu$. Setting $n^*=b\ell^\gamma\asymp \mu$ in Eq.~(\ref{Eq:Sn_1_appendix}), we find
\begin{align}
    \ln{{\ell}\choose{n^*}} & \approx b\ell^\gamma\ln\frac{\ell^{(1-\gamma)}}{b}-\ln\sqrt{2\pi b\ell^\gamma} \nonumber \\
& \quad -(\ell-b\ell^\gamma+\frac{1}{2})\sum_{k=1}^{\infty}\frac{b^k}{k\ell^{(1-\gamma)k}} \label{Eq:Sn_3_appendix}.
\end{align} 
Accordingly,
\begin{align}
\ln{{\ell}\choose{n^*}}\sim b\ell^\gamma\ln\frac{\ell^{(1-\gamma)}}{be}+\mathcal{O}(\ell^{2\gamma-1})
\label{Eq:Sn_4_appendix},
\end{align} 
where the subleading correction is of order $\ell^{2\gamma-1}$ for $1>\gamma>1/2$. For other cases, it exhibits a scaling behavior proportional to the natural logarithm of $\ell$:
\begin{align}
\ln{{\ell}\choose{n^*}}\sim b\ell^\gamma\ln\frac{\ell^{(1-\gamma)}}{be}+\mathcal{O}(\ln \ell)
\label{Eq:Sn_5_appendix}.
\end{align} 
\\
%%%%%%%%%%%%%%%%%%

\section{Prove of inequality Eq.~(\ref{eq:Savg22}) using Jensen’s inequality}
\label{sec:Appendix B}

Here prove the inequality Eq.~(\ref{eq:Savg22}) using Jensen’s inequality.

After rearranging, we can write Eq.~(\ref{eq:SPq_alpha}) as
\begin{align}
    S_{\alpha}=\frac{1}{1-\alpha}\ln\left[\sum_{q_A}P_{q_A} \left(\frac{e^{S_\alpha(q_A)}}{P_{q_A}}\right)^{(1-\alpha)}\right].
\end{align}
For $\alpha>1$, we define the convex function
\begin{align}
    f\left(x \right)=\frac{1}{1-\alpha}\ln\left(x\right),
\end{align}
such that for the positive qualities $X_{q_A}=\left(\frac{e^{S_\alpha(q_A)}}{P_{q_A}}\right)^{(1-\alpha)}$, we can apply Jensen’s inequality for a convex function as 
\begin{align}
    f\left(\sum_{q_A}P_{q_A} X_{q_A} \right)\leq \sum_{q_A}P_{q_A} f\left(X_{q_A}\right).
\end{align}
Therefore
\begin{align}
    S_{\alpha}\leq\sum_{q_A}P_{q_A} \left[S_\alpha(q_A)-\ln{P_{q_A}}\right],
\end{align}
and thus
\begin{align}
S_{\alpha}\leq\sum_{q_A}P_{q_A}S_\alpha(q_A)+H_1(\{P_{q_A}\})\, . 
\end{align}
%
%%%%%%%%%%%%%%%%%%

\printbibliography
\end{document}